\documentstyle[titlepage,12pt]{article}
\begin{document}
\parindent 2em
\baselineskip 4.5ex

\begin{titlepage}
\begin{center}
\vspace{12mm}
{\LARGE Spectral boundary of positive random potential in a strong
magnetic field}
\vspace{25mm}

Igor F. Herbut \\
Department of Physics and Astronomy, Johns Hopkins University, Baltimore,
MD 21218

\end{center}
\vspace{10mm}

\noindent
{\bf Abstract:}
We consider the problem of randomly distributed  positive delta-function
scatterers in a strong magnetic field and study the behavior of
density of states close to the spectral boundary at $E=\hbar\omega_{c}/2$
in both two and three dimensions. Starting from dimensionally reduced
expression of Brezin et al. and using the semiclassical approximation we
show that the density of states in the Lifshitz tail at small energies
is proportional to $e^{f-2}$ in two dimensions and
to $\exp(-3.14 f \ln(3.14 f/\pi e)/\sqrt{2me})$ in three dimensions, where $e$
is the energy and $f$
is the density of scatterers in natural units.


\end{titlepage}

\section{Introduction}
The quantum mechanical problem of particle's
 motion in random external potential
and in strong magnetic field has been a subject of lot of interest in past
years, primarily because of the connection to the problem of quantum
Hall effect. If a particle is confined to two-dimensional (2d) motion
perpendicular to a sufficiently strong
magnetic field the Hilbert space available to it is only
the lowest Landau level (LLL). The average density of states (DOS) has then
been calculated exactly for the white-noise (Gaussian) random potential by
Wegner \cite{1}. Subsequently,
it has been shown that the expression for DOS in the LLL and in
uncorrelated random potential undergoes a spectacular D$\rightarrow$D-2
dimensional reduction \cite{2},
so that Wegner's result can be generalized to any
external potential of this type. For a particle in 3d,
the problem of calculating DOS therefore reduces to solving a 1d
field theory. The exact results are not available in that case,
but the leading behavior of DOS in the tail of the distribution
for the white-noise
potential has been obtained by several authors
using approximate methods \cite{3,4,5}.

 In this paper we study the
spectrum of quantum-mechanical particle
moving in a random distribution of positive $\delta$-function scatterers in
strong magnetic field. The main difference between this and the problem
of Gaussian random potential studied before
is the presence of fixed lower boundary in the spectrum at
$E=\hbar \omega_{c}/2$, where $\omega_{c}$ is the cyclotron energy.
Because of the analytic constraint on the wave-functions
in the LLL spectrum
the structure of DOS close to the boundary is quite
intricate even in 2d case.
The additional interest in this type of random potential
comes from the fact that it can be used to model disorder in studies of
the superconducting ``glassy'' transition in the materials with columnar
or point defects \cite{6}.  The behavior of
DOS and structure of the eigenstates close to the spectral boundary has direct
consequences for the critical behavior in the model. In this paper we
therefore concentrate on DOS close to the spectral boundary
and apply a semiclassical
approximation to the problem to obtain the leading term
\cite{7,8,9}.  Our main results are the following: in 2d
the saddle-point approximation and the direct numerical
diagonalization both confirm the result of ref. 2 that the DOS
is proportional to $e^{f-2}$ where $f$ is the density of scatterers
in appropriate units; in 3d
the leading term in the expression
for logarithm of DOS
is proportional to
$\ln(e)/\sqrt{2me}$ which is
similar but distinct behavior
from purely 1d problem.

The paper is organized in the following manner: in the next
 section we introduce
the basic concepts and a toy-problem: particle in 2d and in the LLL where the
expression for DOS reduces to a simple integral.
In section III we perform the instanton
calculation of the DOS in 3d case and in the last
section we discuss the obtained results.
Finally, in appendix we prove that the obtained
instanton solution is indeed the requisite negative mode of the action.

\section{Two dimensions}

We study the spectrum of the
Hamiltonian:
\begin{equation}
\hat{H}=(-i\hbar \nabla - \vec{A})^2 /2m + \lambda \sum_{i} \delta (
\vec{r}-\vec{r_{i}}),
\end{equation}
where $\vec{A}=(-By/2,Bx/2,0)$,  $\lambda >0$ and $m$ is the mass of the
particle. Coordinates of the scatterers $\{\vec{r_{i}}\}$
are independent random variables and
the average density of scatterers is $\rho$. It is
assumed that the magnetic field
$B$ is strong enough so that the Hilbert space for
the motion orthogonal to the field is
restricted to the LLL. Under that condition, if the particle
is confined to the two dimensions orthogonal to the field, DOS per unit area
is given by the expression \cite{2,10}:
\begin{equation}
\rho(e)=\frac{1}{\pi \lambda} Im \frac{\partial}{\partial e} \ln Z
\end{equation}
where "the average partition function" $Z$ is
\begin{equation}
Z=\int\int_{-\infty}^{+\infty}
 d\phi_{1} d\phi_{2} \exp{(i e (\phi_{1}^2 + \phi_{2}^2) -
f \int_{0}^{\phi_{1}^{2}+\phi_{2}^{2}} \frac{dx}{x}(1-\exp{(-ix)}))}
\end{equation}
and we rescaled the energy as $e=(E-\hbar \omega_{c}/2)2\pi l^{2}/\lambda$,
 $f=\rho 2\pi l^{2}$ and $l$ is the magnetic length.
We assume that $f>1$, since otherwise DOS will have a delta-function
singularity at $e=0$ as a consequence of the nature of the LLL
wave-functions. The fact that $Z$ is
an ordinary integral follows from the hidden supersymmetry in the
problem uncovered by Brezin et al. (ref. \cite{2})
which led to the dimensional reduction by two.
We are interested in the behavior of $\rho(e)$ as $e$
approaches zero from the positive side.
First, we rotate the lines of integration over
variables $\phi_{1}, \phi_{2}$ from the real axes
for $\pi/4$ in the complex plane. The DOS is then given by eq.2 with
\begin{equation}
Z=\int\int d\phi_{1} d\phi_{2} \exp{(-S)},
\end{equation}
and the exponentiated action is
\begin{equation}
S=-e(\phi_{1}^{2}+\phi_{2}^{2})+f \int_{0}^{(\phi_{1}^{2}+\phi_{2}^{2})}
\frac{dx}{x}(1-\exp{(-x)}).
\end{equation}
If $e<0$ the integrand goes to zero
fast when the variables of integration tend to
infinity; the integral is a finite real number and the density of
states vanishes. When $e>0$ however, the integral diverges and we use a
saddle point method to extract the imaginary piece.
Saddle-points of the action $S$ are
determined by $\partial S/\partial \phi_{1/2}=0$, i.e. :
\begin{equation}
\phi_{1/2} (-e + f\frac{1-\exp{(-\phi_{1}^{2}-\phi_{2}^{2}) } }{
\phi_{1}^{2}+\phi_{2}^{2} } ) = 0.
\end{equation}

First, there is a trivial saddle-point $\phi_{1}=\phi_{2}=0$ where the
action vanishes. At this saddle point we have $\partial^{2}S/
\partial \phi_{1/2}^{2}=f-e$ and the mixed derivative is zero. Thus the
contribution of this saddle-point and the quadratic fluctuations around it
to the integral $Z$ is real for $e<f$ and it equals $2\pi/(f-e)$.
 The second set of saddle-points is determined by:
\begin{equation}
\frac{e}{f}(\phi_{1}^{2}+\phi_{2}^{2}) = 1-\exp{(- \phi_{1}^{2} -
\phi_{2}^{2} )}.
\end{equation}
The last equation admits a simple solution in the limit $e<<f$:
$\phi_{1}=\sqrt{f/e}$, $\phi_{2}=0$.
Other solutions are related to this one  by a rotation around the origin in
($\phi_{1},\phi_{2}$) plane. Note that
in the limit of interest ($e<<f$) this saddle point is
infinitely far from the trivial one.
The value of the action at this saddle point
is $S=-.42 f + f \ln (f/e)$, and $\partial^{2}S/\partial \phi_{1}^{2}=-2e$
with all other second derivatives vanishing. Thus, fluctuations in $\phi_{1}$
around this saddle-point represent the "negative mode" and we need to rotate
the line of integration over this variable by $\pi/2$ in the complex plane.
 This
rotation makes the contribution of this saddle-point and fluctuations around
it to the integral purely imaginary. Fluctuations in $\phi_{2}$ represent
the "zero mode" in the problem; the manifestation of breaking of U(1)
symmetry by picking a saddle-point. The integration over
this variable has to be transformed into integration over "collective
coordinate"  with the appropriate Jacobian
\cite{11}, which takes into account the contributions of
 all saddle-points
related by the symmetry. Including the trivial saddle-point, the
result for the partition function when $e/f <<1$ becomes:
\begin{equation}
Z=\frac{2\pi}{f} + i \frac{\exp{(.42 f)}\pi^{3/2}}{f^{f-1/2}} e^{f-1}
\end{equation}
which leads to the result for the DOS when $f>1$:
\begin{equation}
\lambda \rho(e) = \frac{\exp{(.42 f)} (f-1)}{2\pi^{1/2} f^{(f-3/2)}} e^{f-2}.
\end{equation}

This simple analysis yields the correct behavior of DOS at
small energies, and even the coefficient of proportionality is
numerically close to the exact value \cite{2}.
The $e^{f-2}$ dependence is somewhat unexpected
since it is not obvious why the number of states at the boundary should
change from diverging to vanishing at the density of scatterers $f=2$. It would
be interesting to have some intuitive understanding of this feature of
DOS. Also, we note here that the semiclassical analysis of DOS
starting from the full field theory (and not from it's dimensionally reduced
form like it has been done here)
using either the replica trick \cite{8} or the supersymmetry \cite{12}
leads to a wrong power law: $\rho(e) \propto e^{f-1}$.
This comes as a surprise when one recalls that for Gaussian
disorder for instance, both methods give the same behavior
in the tail of DOS as found in the exact solution \cite{8}, \cite{9}.
 We suspect that this is related to the fact that we are not
dealing with the true tail of the distribution here, but
we are close to the fixed edge of the spectrum instead.
Since DOS vanishes only as a power law in our case (in contradistinction
to the Gaussian disorder DOS in the tail) the correct power
could easily be missed by semi-classical treatment. As an independent
check of the validity of the result from dimensionally reduced
expression for DOS we performed numerical diagonalization of the Hamiltonian 1.
On Figure 1 we have shown the result  for DOS obtained
by taking 30 different realizations of the
random potential with degeneracy of the LLL being 100 and density
of scatterers $f=1.5$.
We used the basis of angular momentum eigenstates in the
numerical diagonalization. The abundance of states close to $e=0$
for $1<f<2$ comes from the fact that the LLL wave functions efficiently
 use their zeroes to cover sparse scatterers,
so our choice of the basis is crucial for
revealing the right behavior of DOS when $e\rightarrow 0$.
The result clearly shows that the number of
states at the spectral boundary $e=0$
 still diverges at $f=1.5$ in agreement with the
result 9, although the number of different realizations of the
random potential is too small for a more quantitative comparison.
Numerical diagonalization at $f=2.5$ (Fig. 2) shows that the DOS remains
flat down to the lowest energies, and no divergence is seen at $e=0$. At
energies $e<0.01$ essentially no states are found.

In the above calculation we
assumed that the strengths of all scatterers are the same whilst only their
positions are the random variables. This however is irrelevant for the
obtained power law behavior of DOS close to the boundary;
using the same reasoning as in this section
one can easily show that as long as all strengths
are positive the result $\rho(e)\propto e^{f-2}$ for $e<<f$ holds even if
the strengths
of the scatterers are allowed to fluctuate.

\section{Three dimensions}

We now assume that the particle moves through a full 3d space and that it's
mass is anisotropic with different values along and
orthogonal to the field. Density of states is then given by:
\begin{equation}
\rho(e) = \frac{d}{\pi L \lambda} Im \frac{\partial}{\partial e}
\ln Z
\end{equation}
and the partition function $Z$ is expressed as a functional integral
\begin{equation}
Z=\int D[\phi^{*}(z),\phi(z)] \exp (-S),
\end{equation}
with the action
\begin{equation}
S=\int_{-L/2d}^{L/2d} dz (-e |\phi(z)|^{2} +\frac{|\partial_{z}\phi(z)|^2}
{2m} + f \int _{0}^{|\phi(z)|^{2}} \frac{dx}{x} (1-\exp{(-x)}) ) .
\end{equation}
We chose the unit of length $d=\hbar^{2} 2\pi l^{2}/2\lambda m_{||}$
and $e=(E-\hbar \omega_{c}/2)2\pi l^{2} d/\lambda$,
$m=m_{||}\lambda d/2 \pi l^{2}$ and $f=\rho 2\pi l^{2} d$ are dimensionless.
The cyclotron energy is determined by the mass orthogonal to the field, and
we take the length of the box $L\rightarrow\infty$.
In the limit $m\rightarrow \infty$ the expression for DOS reduces
to its 2d limiting form from the previous section.

Field configuration which minimizes the action is the solution of
the equation:
\begin{equation}
-e\phi(z) - \frac{\partial_{z}^{2}\phi(z)}{2m} +
f\frac{1-\exp{(-|\phi(z)|^{2})}}{|\phi(z)|^{2}} \phi(z) = 0.
\end{equation}
There is again the trivial solution $\phi(z)=0$ which contributes
to the real part of the partition function. To obtain the
imaginary part one needs to find a non-trivial solution
(instanton) of the above equation.
In the region where $\phi^{2}(z)>>f/e$ the instanton is proportional to
$\cos(z\sqrt{2me})$ and where $\phi^{2}(z)<<1$ it decays to zero exponentially.
Instead of solving the nonlinear differential equation 13 we propose
an {\it anzats} for the instanton:
$\phi(z)=a\cos(zb)$
if $ -\pi/2b<z<\pi/2b$ and zero otherwise. The parameters $a$ and $b$ are
to be chosen to minimize the action 12. If we find that
$a^{2}>>f/e$ when $e/f\rightarrow 0$, there will be a wide region where
our anzats will approximate the actual
solution of the above differential equation very well. This variational
procedure is similar to the one used in ref.7 to obtain the tail of DOS
in the same disorder potential but without the magnetic field.

Inserting the proposed anzats into the action we get
\begin{equation}
S= -\frac{e a^{2}\pi}{2b} + \frac{a^{2}b\pi}{4m}
+\frac{f}{b}I_{2}(a),
\end{equation}
and minimizing it with respect to $a^{2}$ and $b$ leads to the
equations:
\begin{equation}
\frac{\pi}{2}(\frac{b^{2}}{2m}-e) +\frac{f}{a^{2}}(\pi - I_{1}(a))=0
\end{equation}
and
\begin{equation}
\frac{a^{2}\pi}{2}(e+\frac{b^{2}}{2m})=f I_{2}(a).
\end{equation}
When $a$ is very large the integrals appearing in the previous
lines can be simplified:
\begin{equation}
I_{1}(a)\equiv\int_{\pi/2}^{\pi/2} dz \exp(-a^{2}\cos^{2}(z))
\approx  \frac{\sqrt{\pi}}{a}
\end{equation}
and
\begin{equation}
I_{2}(a)\equiv\int_{\pi/2}^{\pi/2}dz \int_{0}^{1} \frac{dx}{x} (1-\exp(-xa^{2}
\cos^{2}(z)) )\approx -2.24 + 3.14 \ln(a^{2})
\end{equation}
Eliminating $b$ from the equation 16 leads to the equation for
parameter $a$:
\begin{equation}
a^{2}=\frac{f(3.14 \ln(a^{2})-2.24)}{\pi (e-\frac{f}{a^{2}})},
\end{equation}
which for $e<<f$ has an approximate solution
\begin{equation}
a^{2}\approx\frac{3.14 f}{\pi e} (1-\frac{2.24}{\pi}+ \ln(\frac{3.14 f}{\pi e})
+ \ln(\ln(\frac{3.14f}{\pi e}))).
\end{equation}
Then from equation 15 it follows:
\begin{equation}
b^{2}\approx 2me-\frac{4mf}{a^{2}}.
\end{equation}
The comparison between the anzats determined by these parameters and
numerically determined instanton is shown on Fig. 3. Note that
$a^{2}\propto (f/e)\ln(f/e)$ so that for $e/f\rightarrow0$
the variational parameter indeed
increases faster than $f/e$.

The value of the action at the instanton saddle-point is:
\begin{equation}
S_{0}\approx \frac{3.14 f}{\sqrt{2me}} \ln(\frac{3.14f}{\pi e})
\end{equation}
and we kept only the leading, most diverging term in the limit $e<<f$.
Besides the slower diverging terms which enter $S_{0}$  from
our anzats there will be additional terms coming from the
region where the anzats deviates appreciably from the exact instanton.
These terms can be systematically
investigated starting from the proposed anzats. In appendix we
prove that our variational anzats is indeed
a negative mode of the action, as it is
necessary to get the imaginary part of the partition function.

Since we consider here only the leading term of the logarithm of DOS
when $e<<f$ we ignore the quadratic fluctuations around both
trivial and instanton saddle-point of the action 12. Then from 22, 10
and 11 one obtains
\begin{equation}
\ln \rho(e)=-\frac{3.14 f}{\sqrt{2me}} \ln(\frac{3.14f}{\pi e})+... .
\end{equation}
The leading behavior is similar as in the
corresponding purely 1d problem where one
obtains the Lifshitz tail $\ln \rho(e)\propto -1/\sqrt{2me}$
when $e\rightarrow 0$.
Note however that DOS in our case
vanishes faster at small energies than in 1d.

\section{Discussion}

In 2d the behavior of DOS in the limit of small energies
$e/f<<1$ depends critically on the density of scatterers: it diverges
for $1<f<2$, goes to constant when $f=2$
 and to zero when $2<f$.  In contrast, in 3d
DOS resembles more the familiar 1d case; irrespectively of $f$
DOS vanishes exponentially fast in the limit $e/f\rightarrow 0$. One
expects that the states at the bottom of the band in 3d are
localized in the rare large regions free of impurities, which would
roughly correspond to the found behavior of DOS \cite{13}.
To quantitatively study the relation between 3d DOS in the
limit when $m\rightarrow \infty$ (i. e. when the mass parallel to the field
becomes very large) and the 2d result
one needs to include the fluctuations around the
saddle points studied in the previous section and the next order terms
in the instanton action 22. We will not dwell on this
here, since it is already possible to see
qualitatively what happens.
Ignoring the energy dependence that comes from the quadratic fluctuations,
 we may write
$Z\approx 1+i \exp(-S_{0})$, where the first term is the contribution
from the trivial saddle-point and the imaginary peace comes from the
instanton. Differentiating the logarithm of $Z$
with respect to the energy and taking the
imaginary peace leads to DOS: $\rho(e)\propto ln(1/e)\exp(-S_{0})
/\sqrt{2me^{3}}$. At fixed density of scatterers $f$, the energy at which
DOS reaches its maximum value goes to zero
when $m\rightarrow\infty$ as illustrated on Fig 4.
On the other hand, when $m$ is
constant increasing the density of scatterers $f$ makes the peak of DOS
flatter (Fig. 5). Thus, having a finite (as opposed to infinite)
mass along the field basically
shifts the maximum of DOS from
being at $e\approx 0$ for $f\approx 2$ to reside at some
finite energy $e_{max}\propto 1/m$. The behavior of DOS close and right to
the maximum resembles its 2d counterpart, while left to it DOS drops
sharply to zero, so that for $e<<1/m$ there are essentially no states.

The described behavior resembles very much the situation in 1d
in the limit of weak disorder \cite{13,14}. In that case DOS differs
from the one in an ideal system only in the narrow region between zero
and $c^2$, where $c=\rho/\lambda$ is the parameter characterizing the
strength of disorder.

\section{Acknowledgments}

It is a pleasure to thank Professor Zlatko Te\v sanovi\' c for many
valuable discussions. We are also grateful to Seunghun Lee for
 his generous technical help.

\section{Appendix}

For the partition function 11 to have an imaginary part it is essential
that the instanton saddle-point discussed in section 3 has a mode
with a negative eigenvalue. The rotation of the line of integration
over this mode by $\pi/2$ in the complex plane then makes the
contribution of this saddle-point to the integral purely imaginary.
Here we show that the proposed anzats is indeed such a mode.  To study
fluctuations around the non-trivial saddle-point configuration
$\phi_{0}(z)$ one needs to diagonalize the operator
\begin{equation}
\hat{O}=\frac{\delta^{2}S}{\delta\phi(z')\delta\phi^{*}(z)}|_{\phi_{0}} =
-\frac{\partial_{z}^{2}}{2m}-(e-f\exp(-|\phi_{0}(z)|^{2}))\delta(z-z').
\end{equation}
We take  the operator at $\phi_{0}(z)=a\cos(zb)$ with $a$ and $b$
given as in eqs. 20 and 21 and calculate the matrix element
$\langle \phi_{0}|\hat{O}|\phi_{0}\rangle$. In the limit of large
$a$ it is straightforward to obtain:
\begin{equation}
\langle\phi_{0}|\hat{O}|\phi_{0}\rangle=-\frac{f\pi}{b} (1-\frac{1}{2a
\sqrt{\pi}})
\end{equation}
In the limit $e/f\rightarrow 0$ we have $a\rightarrow \infty$ so from
the last equation it follows that in the same limit
$\langle\phi_{0}|\hat{O}|\phi_{0}\rangle
 < 0$. Thus our anzats has a non-zero overlap with the exact negative mode
of the operator $\hat{O}$ taken at $\phi_{0}$. This is sufficient to
make the contribution of the corresponding saddle-point to the partition
function purely imaginary.

\pagebreak
Captions:

Figure 1. DOS in 2d at $f=1.5$ obtained by taking 30 different
realizations of random potential in LLL with degeneracy 100. In the
inset DOS close to zero energy is shown.

Figure 2. Same is in Figure 1 but at f=2.5 and for 10 realizations
of random potential.

Figure 3. Numerical solution of eq. 13 (dots) and the anzats
determined by the parameters from eq. 20 and 21 (full line) at
energy $e/f =10^{-4}$ and $2m=1$.

Figure 4. Approximate expression for DOS $\rho(e)\approx \ln(1/e)
\exp(-S_{0})/\sqrt{2me^{3}}$ (see the text) at $f=1.5$ for three values of the
mass $m=1000,2000,4000$. The peak shifts to the left as the mass
increases.

Figure 5. The same expression for DOS as in figure 3. but with the mass
fixed at $m=1000$ and density of scatterers varied: f=1.5,2,2.5 from
top to bottom.

\pagebreak

\end{document}